\documentclass{aa}

\usepackage{txfonts}
\usepackage{graphicx}
\usepackage{natbib}
\bibpunct{(}{)}{;}{a}{}{,}

\long\def\symbolfootnote[#1]#2{\begingroup\def\thefootnote{\fnsymbol{footnote}}\footnote[#1]{#2}\endgroup}

\begin{document}

\title{Detailed study of the GRB\,030329 radio afterglow deep into the non-relativistic phase}

\author{A.J.~van~der~Horst\inst{1} 
\and A.~Kamble\inst{2} 
\and L.~Resmi\inst{2,3} 
\and R.A.M.J.~Wijers\inst{1} 
\and D.~Bhattacharya\inst{2,4} 
\and B.~Scheers\inst{1} 
\and E.~Rol\inst{5} 
\and R.~Strom\inst{6,1} 
\and C.~Kouveliotou\inst{7} 
\and T.~Oosterloo\inst{6} 
\and C.H.~Ishwara-Chandra\inst{8}}

\offprints{A.J. van der Horst, \email{avdhorst@science.uva.nl}}

\institute{Astronomical Institute, University of Amsterdam, Kruislaan 403, 1098 SJ Amsterdam, The Netherlands
\and Raman Research Institute, Bangalore 560080, India
\and Joint Astronomy Programme, Indian Institute of Science, Bangalore 560012, India
\and Inter-University Centre for Astronomy and Astrophysics, Pune 411007, India
\and Department of Physics and Astronomy, University of Leicester, University Road, Leicester LE2 7RH, UK
\and ASTRON, P.O. Box 2, 7990 AA Dwingeloo, The Netherlands
\and NASA/MSFC, NSSTC, VP62, 320 Sparkman Drive, Huntsville, AL 35805, USA.
\and National Centre for Radio Astrophysics, Post Bag 3, Ganeshkind, Pune 411007, India}



\abstract
{We explore the physics behind one of the brightest radio afterglows ever, GRB\,030329, at late times when the jet is non-relativistic.}
{We determine the physical parameters of the blast wave and its surroundings, in particular the index of the electron energy distribution, 
the energy of the blast wave, and the density (structure) of the circumburst medium. 
We then compare our results with those from image size measurements.}
{We observed the GRB\,030329 radio afterglow with the Westerbork Synthesis Radio Telescope and the Giant Metrewave Radio Telescope 
at frequencies from 325~MHz to 8.4~GHz, spanning a time range of 268-1128~days after the burst. 
We modeled all the available radio data and derived the physical parameters.}
{The index of the electron energy distribution is $p=2.1$, the circumburst medium is homogeneous, 
and the transition to the non-relativistic phase happens at $t_{\rm{NR}}\sim 80$~days. 
The energy of the blast wave and density of the surrounding medium are comparable to previous findings.}
{Our findings indicate that the blast wave is roughly spherical at $t_{\rm{NR}}$, 
and they agree with the implications from the VLBI studies of image size evolution. 
It is not clear from the presented dataset whether we have seen emission from the counter jet or not. 
We predict that the Low Frequency Array will be able to observe the afterglow of GRB\,030329 and many other radio afterglows, 
constraining the physics of the blast wave during its non-relativistic phase even further.}


\titlerunning{GRB\,030329 radio afterglow deep into the non-relativistic phase}
\authorrunning{Van der Horst et al.}

\maketitle


\section{Introduction}\label{section:introduction}

GRB\,030329 has been a very distinctive event in many respects. 
Residing at a redshift of 0.1685 \citep{greiner2003:gcn2020}, i.e. at a luminosity distance of 802~Mpc 
(adopting a flat universe with $\Omega_{\rm{M}} = 0.27$, $\Omega_{\Lambda} = 0.73$ and $\rm{H}_{0} = 71 \rm{km} \rm{s}^{-1} \rm{Mpc}^{-1}$), 
it is one of the nearest GRBs for which an afterglow has been observed (GRB\,980425 at $\rm{z} = 0.0085$ remains the nearest of the GRBs). 
GRB\,030329 displayed one of the brightest afterglows ever, enabling the study of its evolution for a long time and in detail over a broad range of frequencies, 
from X-ray to centimetre wavelengths. The afterglow of this GRB is one with the longest follow ups ever, still visible in radio waves three years after the burst trigger. 
It was also the first GRB to have provided unambiguous evidence of the long suspected \citep[e.g.][]{galama1998:nature395} 
association between GRBs and supernovae \citep{hjorth2003:nature423,stanek2003:apj591}. 

GRB\,030329 was detected and localised by the HETE-II satellite \citep{vanderspek2003:gcn1997} on 29th March 2003, at UT 11:37:14.7 and lasted more than 100~s. 
The measured fluence for this burst was $5.5 \times 10^{-5} \rm{erg}\,\rm{cm}^{-2}$ in the 7-30~keV band, 
and $1.1 \times 10^{-4} \rm{erg}\,\rm{cm}^{-2}$ in the 30-400~keV band.
The burst was followed by an extremely bright X-ray afterglow, $1.4 \times 10^{-10} \rm{erg}\,\rm{cm}^{-2}\,\rm{s}^{-1}$ in the 2-10~keV band, 
detected by RXTE $\sim 5$~h after the burst \citep{marshall2003:gcn1996}. 
The optical afterglow was detected, 67 minutes after the burst, in R band at $12.4$~mag \citep{sato2003:apj599}. 
A bright radio afterglow of 3.5~mJy at 8.46~GHz was detected by the Very Large Array (VLA) on 2003 March 30.06 UT \citep{berger2003:nature426}. 
Around 7 days after the burst the optical spectrum showed the signature of underlying supernova emission, 
SN~2003dh \citep{hjorth2003:nature423,stanek2003:apj591}. 
The afterglow was subsequently followed at X-ray, optical, millimeter and radio wavelengths, providing the extremely rich temporal coverage of the transient 
in all the wavebands \citep[e.g.][]{tiengo2004:aa423,lipkin2004:apj606,gorosabel2006:apj641,sheth2003:apj595,kuno2004:pasj56,berger2003:nature426}. 
The proximity of the GRB and the exceptional radio brightness of its afterglow made it possible to resolve the afterglow image 
with a Very Long Baseline Interferometry (VLBI) campaign from 8~days \citep{taylor2004:apj609} up to 806~days \citep{pihlstrom2007:astroph07042085} after the burst. 

Extensive radio follow up of the afterglow has been reported earlier by 
\citep{berger2003:nature426}, \citet{frail2005:apj619}, \citet{vanderhorst2005:apj634} and \citet{resmi2005:aa440}. 
These reports cover a period of up to 1 year post burst.
In this paper, we report further extension of the low frequency radio follow up of the GRB\,030329 afterglow, up to 1128~days after the burst,
using the Westerbork Synthesis Radio Telescope (WSRT) and the Giant Metrewave Radio Telescope (GMRT). 
GRB\,030329 is the first afterglow to be detected below an observing frequency of 1~GHz, and even as low as 610~MHz. 
In Section \ref{section:observations} we describe the observations and data analysis. 
In Section \ref{section:results} we present the results from modeling the light curves of all the available radio data of the GRB\,030329 afterglow, 
focusing on the blast wave physics in the non-relativistic phase of its evolution. 
In Section \ref{section:discussion} we compare our modeling results with previous light curve studies and with the VLBI image size evolution. 
Furthermore, we put constraints on the emission of the counter jet, and we show model predictions for radio afterglow observations with the Low Frequency Array (LOFAR). 
Section \ref{section:conclusions} summarises our results.


\section{Observations \& Analysis}\label{section:observations}

GRB\,030329 was observed with the WSRT and GMRT from 325~MHz to 8.46~GHz, spanning a time range of 268-1128~days after the burst. 
The afterglow was clearly detected at all frequencies except for 325~MHz, where we only obtained upper limits. 
GRB\,030329 is the first afterglow to be detected at frequencies less than 1~GHz: at 840~GHz with WSRT and even as low as 610~MHz with GMRT. 
Here we describe the data reduction and analysis of our observations; the observational results are summarised in Table~\ref{table:wsrtresults} 
and Table~\ref{table:gmrtresults}, for WSRT and GMRT respectively.


\subsection{WSRT Observations}

The first observations of GRB\,030329 with WSRT were carried out at $\sim 1.4$ days after the burst, at 1.4, 2.3 and 4.8~GHz. 
After the clear detections of the afterglow, we started an intensive monitoring campaign at these three frequencies, 
of which the first results were presented in \citet{vanderhorst2005:apj634}. 
Here we present the results of follow up measurements up to 1128~days after the burst in a wider frequency range, 
adding detections at 840~MHz and 8.4 GHz, and upper limits at 350 MHz. 
We used the Multi Frequency Front Ends \citep{tan1991:aspcs19} in combination with the IVC+DZB backend\footnote{See Section 5.2 at 
http://www.astron.nl/wsrt/wsrtGuide/node6.html} in continuum mode, with a bandwidth of 8x20 MHz. 
Gain and phase calibrations were performed with the calibrator 3C286, though sometimes 3C147 or 3C48 were used. 
The observations were analysed using the Multichannel Image Reconstruction Image Analysis and Display (MIRIAD) software package; 
except for the 350 MHz observations, which were analysed using the Astronomical Image Processing System (AIPS). 
Table \ref{table:wsrtresults} lists the log of the observations, spanning a time range of 268-1128 days after the burst. 
We checked our results for consistency by measuring the flux of several nearby point sources in the primary beam, which were assumed to be constant in time. 
We found no significant correlated flux-variations between the various epochs.

\begin{table*}
\caption{Log of WSRT observations of GRB\,030329.}
\label{table:wsrtresults}
\centering
\begin{minipage}{\textwidth}
\centering
\begin{tabular}{l c c c c}
\hline\hline
Observing Dates & $\Delta$T\symbolfootnote[1]{The indicated time is the logarithmic average of the start and end of the integration.} & Integration time & Frequency & Flux\symbolfootnote[2]{The measurement uncertainties are given at $1 \sigma$ level.} \\
  & (days) & (hours) & (GHz) & ($\mu$Jy) \\
\hline
2003 Dec 22.926 - 23.196 &  268.577 &  6.5 & 2.3 & $1618\pm  31$ \\
2003 Dec 25.005 - 25.188 &  270.612 &  4.4 & 1.4 & $2502\pm 139$ \\
2004 Jan 25.833 - 26.333 &  302.599 & 12.0 & 0.35 & $<951\,(3\sigma)$\symbolfootnote[3]{A meaningful formal flux measurement at the GRB position cannot be determined because of confusion by a very nearby, bright source.} \\
2004 Jan 29.822 - 29.961 &  306.408 &  3.3 & 1.4 & $1552\pm 111$ \\
2004 Jan 29.995 - 30.134 &  306.580 &  3.3 & 4.8 & $<648\,(3\sigma)$\symbolfootnote[4]{Formal flux measurement at the GRB position gives $1\pm 216$ $\mu$Jy.} \\
2004 Jan 30.168 - 30.307 &  306.753 &  3.3 & 2.3 & $1389\pm  67$ \\
2004 Jan 31.012 - 31.137 &  307.591 &  3.0 & 8.4 & $ 815\pm 225$ \\
2004 Feb 10.790 - 11.289 &  318.555 & 12.0 & 0.84 & $2332\pm 288$ \\
2004 Mar 26.667 - 26.934 &  363.816 &  6.4 & 4.8 & $ 597\pm  27$ \\
2004 Mar 27.664 - 28.152 &  364.424 & 11.7 & 1.4 & $1318\pm 104$ \\
2004 Mar 28.862 - 29.153 &  365.524 &  7.0 & 2.3 & $ 871\pm  29$ \\
2004 Apr 11.623 - 11.880 &  379.267 &  6.2 & 0.84 & $1525\pm 389$ \\
2004 May  2.566 -  3.065 &  400.331 & 12.0 & 0.35 & $<1305\,(3\sigma)^c$ \\
2004 May 19.634 - 20.018 &  417.342 &  9.2 & 1.4 & $1824\pm 100$ \\
2004 May 22.511 - 22.794 &  420.168 &  6.8 & 2.3 & $ 933\pm  34$ \\
2004 Jul  3.396 -  3.729 &  462.078 &  8.0 & 2.3 & $ 707\pm  39$ \\
2004 Jul  4.394 -  4.726 &  463.076 &  8.0 & 4.8 & $ 329\pm  27$ \\
2004 Aug  1.317 -  1.598 &  490.974 &  6.7 & 1.4 & $ 622\pm  95$ \\
2004 Sep 25.328 - 25.667 &  546.013 &  8.1 & 4.8 & $ 274\pm  34$ \\
2004 Nov  2.063 -  2.271 &  583.683 &  5.0 & 2.3 & $ 543\pm  46$ \\
2004 Nov 11.039 - 11.371 &  592.721 &  8.0 & 1.4 & $1162\pm  63$ \\
2004 Nov 12.036 - 12.536 &  593.802 & 12.0 & 0.84 & $1306\pm 366$ \\
2004 Nov 20.014 - 20.486 &  601.766 & 11.3 & 0.35 & $<2124\,(3\sigma)^c$ \\
2005 Mar 24.673 - 25.172 &  726.439 & 12.0 & 0.84 & $1199\pm 238$ \\
2005 Mar 26.667 - 27.167 &  728.433 & 12.0 & 0.35 & $<996\,(3\sigma)^c$ \\
2005 Apr  9.629 - 10.129 &  742.395 & 12.0 & 1.4 & $1078\pm  72$ \\
2005 Apr 10.627 - 11.126 &  743.892 & 12.0 & 2.3 & $ 504\pm  48$ \\
2005 May 14.534 - 14.919 &  777.242 &  9.2 & 4.8 & $ 409\pm  27$ \\
2005 May 15.531 - 15.883 &  778.223 &  8.1 & 8.4 & $<309\,(3\sigma)$\symbolfootnote[5]{Formal flux measurement at the GRB position gives $-69\pm 103$ $\mu$Jy.} \\
2005 Nov 27.993 - 28.493 &  974.759 & 12.0 & 4.8 & $ 150\pm  19$ \\
2005 Dec  7.004 -  8.292 &  984.163 & 13.8 & 2.3 & $ 318\pm  58$ \\
2005 Dec  9.000 - 10.233 &  986.132 & 12.5 & 1.4 & $ 842\pm 127$ \\
2006 Apr  9.637 - 10.129 & 1107.399 & 11.8 & 0.84 & $ 817\pm 392$ \\ 
2006 Apr 30.573 - May  1.072 & 1128.338 & 12.0 & 4.8 & $ 157\pm  22$ \\ 
\hline
\end{tabular}
\end{minipage}
\end{table*}


\subsection{GMRT Observations}

The radio afterglow of GRB 030329 was first detected by GMRT at 1280~MHz on the 31st of March 2003, 2.3~days after the burst \citep{rao2003:gcn2073} with a flux of 0.25~mJy. 
The afterglow was observed since then at 1280~MHz, 610~MHz and 325~MHz. 
We observed the afterglow at a total of 27 epochs (12 epochs at 1280~MHz, 13 epochs at 610~MHz and 2 epochs at 325~MHz), 
excluding the first year observations reported in \citet{resmi2005:aa440} (9 epochs at 1280 MHz).
We have used a bandwidth of 32~MHz for all these observations.
One of the three possible flux calibrators, 3C48, 3C147 or 3C286, was observed at the beginning and end of each observing session for about 15 minutes, 
as a primary flux calibrator to which the flux scale was set. 
Radio sources 1125+261 and 1021+219 were used as phase calibrators at 1280~MHz and 610~MHz, respectively.
The phase calibrator was observed for about 6~minutes before and after an observing scan of about 30 to 45~minutes on GRB\,030329.
The data thus recorded were then converted to FITS format and analysed using AIPS. 
Fluxes of the individual sources were measured using the task `jmfit' in AIPS.

During the data reduction, we had to discard a fraction of the data affected by radio frequency interference and system malfunctions. 
At the end of this we could use an effective bandwidth of about 13~MHz, on average, for all the observations reported here. 
Estimated fluxes of other sources in the field of GRB\,030329, i.e. in the primary beam, showed a variation of up to $\sim$20~\% between observations at different  epochs. 
A part of this variation appears to be correlated between different sources. 
This indicates the presence of possible systematic uncertainties in the flux calibration arising from instrumental effects, for which we attempt to correct in the following manner. 
At each frequency band, we have chosen a reference observation: at 610~MHz we have chosen the field observed on September 2 2005 as a reference, 
and at 1280~MHz the one observed on July 1 2005. 
For every other observation, we selected four sources in the field within $5\arcmin$ of the GRB\,030329 position and estimated their fluxes. 
Ratios of these estimated fluxes to those at the reference epoch were computed and averaged for each observation. 
The flux of the afterglow measured at each epoch was then scaled using this average flux ratio. 
The correction to the GRB flux obtained from this procedure was quite small, typically less than the map r.m.s. at most epochs, except at two instances at 1280~MHz where the correction amounted to $\sim 2\sigma$. 
This gives us confidence that the estimated flux of the GRB is quite secure at both 610 and 1280~MHz. 
The final fluxes, with the corrections mentioned above incorporated, are presented in Table~\ref{table:gmrtresults}.

\begin{table*}
\caption{Log of GMRT observations of GRB\,030329.}
\label{table:gmrtresults}
\centering
\begin{minipage}{\textwidth}
\centering
\begin{tabular}{l c c c}
\hline\hline
Observing Dates & $\Delta$T & Frequency & Flux\symbolfootnote[2]{The measurement uncertainties are given at $1 \sigma$ level.} \\
                & (days)    & (MHz)     & (mJy) \\
\hline
2004 Jan 4.97   &  281.49  &   610  &  $0.90\pm 0.20$ \\
2004 Feb 26.63  &  334.15  &   610  &  $1.14\pm 0.25$ \\
2004 Apr 10.49  &  378.01  &   610  &  $0.61\pm 0.11$ \\
2004 Jun 27.28  &  455.80  &   610  &  $1.01\pm 0.18$ \\
2004 Jun 28.28  &  456.80  &  1280  &  $1.11\pm 0.13$ \\
2004 Jul 18.31  &  476.83  &  1280  &  $1.04\pm 0.11$ \\
2004 Jul 20.28  &  478.80  &   610  &  $0.84\pm 0.17$ \\
2004 Aug 20.04  &  509.56  &  1280  &  $1.03\pm 0.11$ \\
2004 Sep 06.00  &  526.52  &   610  &  $0.98\pm 0.15$ \\
2004 Sep 17.96  &  538.48  &  1280  &  $0.48\pm 0.10$ \\
2004 Dec 3.00   &  614.52  &   610  &  $1.08\pm 0.35$ \\
2005 Jan 6.82   &  649.34  &   610  &  $0.50\pm 0.11$ \\
2005 Feb 8.81   &  682.33  &   325  &  $<1.5 (3\sigma)$\symbolfootnote[2]{A meaningful formal flux measurement at the GRB position cannot be determined because of confusion by a very nearby, bright source.} \\
2005 Feb 13.84  &  687.34  &  1280  &  $0.97\pm 0.14$ \\
2005 Feb 21.71  &  695.23  &   610  &  $0.47\pm 0.13$ \\
2005 Mar 10.77  &  712.29  &  1280  &  $0.72\pm 0.07$ \\
2005 Mar 18.66  &  720.18  &   610  &  $0.69\pm 0.13$ \\
2005 Jun 17.54  &  811.06  &   610  &  $<1.1 (3\sigma)$\symbolfootnote[3]{Formal flux measurement at the GRB position gives $0.74\pm 0.36$ mJy.} \\
2005 Jun 28.59  &  702.11  &   325  &  $<1.8 (3\sigma)^b$ \\
2005 Jul 1.19   &  824.71  &  1280  &  $0.53\pm 0.09$ \\
2005 Sep 2.99   &  888.51  &   610  &  $0.68\pm 0.13$ \\
2005 Oct 8.20   &  923.72  &  1280  &  $0.49\pm 0.09$ \\
\hline
\end{tabular}
\end{minipage}
\end{table*}


\section{Modeling Results}\label{section:results}

\begin{figure*}
  \centering
  \includegraphics[width=\textwidth]{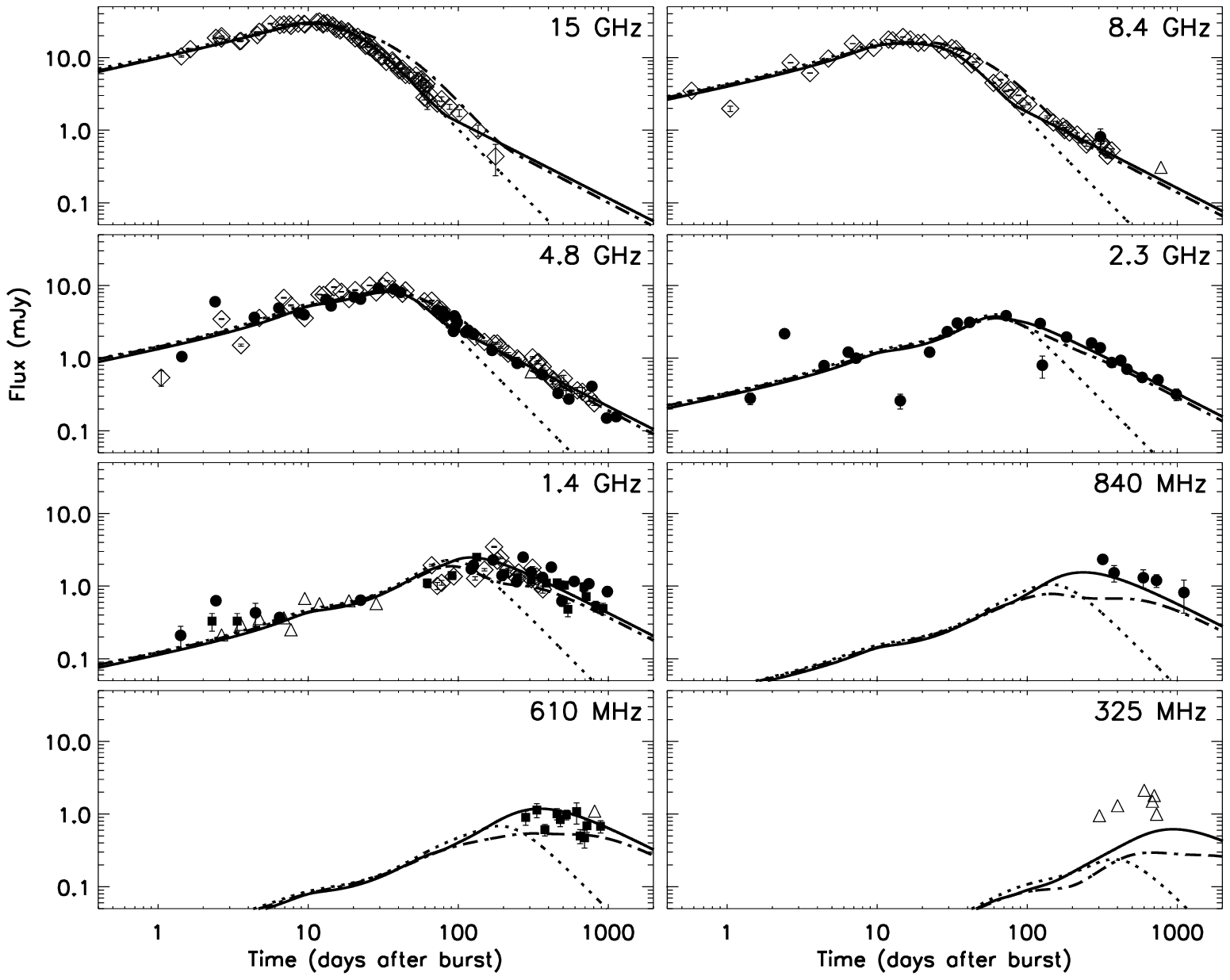}
  \caption{Modeling results of the afterglow of GRB\,030329 at centimetre wavelengths. 
    Our light curves obtained with WSRT and GMRT are shown together with previously reported fluxes from WSRT \citep{vanderhorst2005:apj634}, GMRT \citep{resmi2005:aa440}, 
    and VLA, ATCA \& Ryle Telescope \citep{berger2003:nature426,frail2005:apj619}. 
    The filled circles and squares are WSRT and GMRT measurements respectively; the open diamonds are VLA, ATCA \& Ryle Telescope measurements; the open triangles are $3 \sigma$ upper limits. 
    Three fits to the data are shown: the dotted line represents a fit to the first 100 days of radio observations with a wide jet expanding in a homogeneous medium; 
    the solid line corresponds to a model in which the blast wave becomes non-relativistic after 80 days; 
    the dash-dotted line corresponds to a model in which a third jet-component with an even wider opening angle is present. 
    The latter model is excluded by the observations below 1~GHz, 
    which leaves the model with the non-relativistic phase after 80 days as the preferred model for the late-time behaviour of the blast wave.}
  \label{fig:results}
\end{figure*}

We have modeled the light curves obtained together with previously reported fluxes from WSRT \citep{vanderhorst2005:apj634}, GMRT \citep{resmi2005:aa440}, 
and VLA, ATCA \& Ryle Telescope \citep{berger2003:nature426,frail2005:apj619,pihlstrom2007:astroph07042085}, see Figure \ref{fig:results}. 
The light curves show the characteristics that are expected for the low-frequency part of a GRB afterglow. 
Since both the peak frequency $\nu_{\rm{m}}$ of the spectral energy distribution and the synchrotron self-absorption frequency $\nu_{\rm{a}}$ 
are situated above the radio regime at early times and they both decrease in time, the light curves rise. 
When the peak of the broadband synchrotron spectrum, either $\nu_{\rm{m}}$ or $\nu_{\rm{a}}$, has moved through the observing band, 
the light curves turn over and decline steeply. 

According to the blast wave model \citep{rees1992:mnras258,meszaros1997:apj476,wijers1997:mnras288} the afterglow of a GRB 
is due to non-thermal synchrotron radiation emitted by shock accelerated electrons.
The large amount of energy released during the burst, with a collimated outflow to start with, drives a powerful relativistic blast wave. 
As the blast wave propagates into the circumburst medium, 
the electrons in the medium are accelerated to a power-law distribution of Lorentz factors with power-law index $p$. 
These relativistic electrons gyrate in the post-shock magnetic field and emit synchrotron radiation which is seen as the afterglow of the GRB. 
The power-law distribution of the electrons results in a power-law spectrum of the afterglow. 
Meanwhile the shock wave decelerates as it propagates into the circumburst medium. 
Assuming mass-energy conservation across the shock front, it has been shown that the Lorentz factor of the blast wave falls off as a power-law with its radius. 

It is expected that the decelerating shock wave becomes non-relativistic after a few weeks. 
Somewhat earlier, the sideways expansion of the initially tightly collimated outflow would become important, 
and by the time of the non-relativistic transition the shock wave becomes nearly spherical. 
In the non-relativistic phase the evolution of the shock wave can be described using the Sedov-Taylor self similar solutions. 
A detailed description of this phase can be found in \citet{frail2000:apj537} (hereafter FWK00). 
Observations of the broadband afterglow during this phase can be used to estimate various physical parameters related to the explosion, 
such as the amount of energy released during the explosion, fractional amount of energy in the accelerated electrons and in the post-shock magnetic field, 
and the density and structure of the circumburst medium. 
These parameters can also be estimated, independently, by modeling the evolution of the afterglow in the relativistic phase, 
but this suffers from uncertainties related to collimation geometry and relative orientation of the observer, problems that do not plague the non-relativistic phase.

The broadband afterglow of GRB\,030329, from radio to X-ray frequencies, can be modeled by the standard relativistic blast wave model, 
assuming either a `double jet model' \citep{berger2003:nature426} or a `refreshed jet model' \citep{resmi2005:aa440}. 
In both the models, the early-time optical and X-ray light curves are explained by a jet with a small opening angle of $\sim 5^{\circ}$. 
The double jet model assumes a co-aligned wider jet component ($\sim 20^{\circ}$) that carries the bulk of the energy and produces the later time light curves. 
In the refreshed jet model, the initial jet is re-energised by the central engine during its lateral expansion, re-collimating it to a wider opening angle. 
This refreshed jet then produces the late-time emission. Unfortunately, we cannot distinguish between the two models with all the available data, 
including the data presented in this paper.

After 80 days the observed radio light curves flatten, which can be explained by a transition 
into the non-relativistic phase of the blast wave \citep{vanderhorst2005:apj634,resmi2005:aa440,frail2005:apj619}. 
It was suggested in \citet{vanderhorst2005:apj634} that this late-time behaviour could also be explained by a third jet-component 
with an even wider opening angle than the first two. 
The latter model, however, is excluded by the observations below 1~GHz (see Figure \ref{fig:results}), 
which leaves the model with the non-relativistic phase after 80 days as the preferred model for the late-time behaviour of the blast wave. 
The precise value of the time of the transition into the non-relativistic phase varies between 60 to 80 days 
because of the different modeling methods applied by various authors. 
The transition into the non-relativistic phase is also observed at X-ray frequencies by \citet{tiengo2004:aa423}, who estimate a transition time of $\sim 44$~days. 
The X-ray light curve, however, is less well sampled than the radio light curves at centimetre wavelengths, which makes the determination of the transition time uncertain. 

Because most of the GRBs occur at cosmological distances, they are not bright enough to be observed at late times, 
and in fact most of them fade below the detection limits much before the start of the non-relativistic phase.
GRB\,030329, being one of the closest GRBs, provides us with a unique example of late time observations deep into the non-relativistic phase.
Since the first months of the radio afterglow have been extensively discussed in the literature, we will focus on the late-time behaviour of the light curves. 
Here we present a detailed analysis of the non-relativistic phase of the afterglow.


\subsection{Spectral \& Temporal Slopes}\label{section:slopes}

We investigate the spectral and temporal slopes in the non-relativistic phase by performing a joint temporal 
and spectral power-law fit of $F_{\nu} = F_{\rm{n}}\cdot (\nu_{\rm{GHz}}/4.86)^{-\beta}\cdot (t_{\rm{days}}/100)^{-\alpha}$. 
Because of the uncertainty in the exact transition time into the non-relativistic phase, $t_{\rm{NR}}$, 
and because the value for the temporal slope could be altered by the smoothness of the transition, 
we use all the available data after 100~days at 2.3, 4.8, 8.4, 15 and 22.5~GHz. 
At 1.28 \& 1.4~GHz we use all the data after 300~days, as the light curve at these frequencies peaks around 100~days. 
We find $\alpha = 1.08\pm 0.03$, $\beta = 0.54\pm 0.02$ and $F_{\rm{n}} = 2.69\pm 0.06$~mJy. 

After 100~days the peak frequency and synchrotron self-absorption frequency have passed through the observing bands we used for this fit, 
and we do not find any evidence for a chromatic break, caused by the so-called cooling break, up to 1128~days. 
So in this case the spectral slope $\beta$ is given by $(p-1)/2$, and the temporal slope $\alpha$ is given by $3(5p-7)/10$ or $(7p-5)/6$, 
for a homogeneous and wind circumburst medium respectively. The power-law index $p$ of the electron energy distribution is thus given by $2\beta+1$, 
and $(10\alpha+21)/15$ (homogeneous medium) or $(6\alpha+5)/7$ (wind medium). For our fitted values we find $p = 2.09\pm 0.03$ from the spectral slope, 
and $p = 2.12\pm 0.02$ (homogeneous) or $p = 1.64\pm 0.02$ (wind) from the temporal slope. 
This indicates that the circumburst medium is homogeneous in the non-relativistic phase of the blast wave evolution, 
and that a wind medium is completely rejected in this phase. 
If one assumes that the circumburst medium density is a power-law of the radius $r$ with index $k$, i.e. proportional to $r^{-k}$, 
the value of $k$ and its uncertainty can actually be calculated directly from $\alpha$ and $\beta$: $k = (5\alpha-15\beta+3) / (\alpha-4\beta+2)$. 
For our fitted values we get $k = 0.33^{+0.34}_{-0.41}$, consistent with a homogeneous medium ($k=0$); 
and this value is inconsistent with a wind medium ($k=2$) at the $5 \sigma$ level. 

The value of $p=2.1$ is in agreement with values found by other authors, although it does matter which data one includes, 
i.e. whether one takes only the data after 100~days or also earlier data. 
Including earlier data gives higher values for $\alpha$ and $\beta$, and as a result a higher value for $p$, with a significantly higher reduced chi-squared. 
This indicates that if one includes part of the data before 100~days, a steeper decay phase is also sampled, i.e. the jet-spreading phase. 
And thus our choice of fitting the data after 100~days is a valid assumption.


\subsection{Physical Parameters}\label{section:parameters}

We have used 1128 days of broadband radio observations (610 MHz to 43.3 GHz) of the afterglow of GRB\,030329 to model the dynamical evolution 
of the afterglow as well as to constrain the explosion energy. 
In Figure \ref{fig:results} we compare our model predictions with the observations. 
The model presented in \citet{vanderhorst2005:apj634}, which fitted the data up to eight months, 
perfectly fits the data up to more than three years after the burst. 
In \citet{vanderhorst2005:apj634} values for the peak frequency $\nu_{\rm{m}}$ of 35 GHz and for the self-absorption frequency $\nu_{\rm{a}}$ of 13 GHz 
were found, both measured at the jet-break time of 10~days; the flux at $\nu_{\rm{m}}$ at that time was 61~mJy. 
After about 80 days the afterglow shows a flattening of the light curve, which is the start of the non-relativistic phase of the explosion: $t_{\rm{NR}}$. 

The fitted spectral parameters may now be used to derive the physical parameters of the explosion. 
The break frequencies and the peak flux, estimated deep in the non-relativistic phase, i.e. at a reference time $t_0 \gg t_{\rm NR}$, 
can be used to yield an estimate of the blast wave energy $E_{\rm ST}$ and the ambient baryon density $n_i$, 
using the Sedov-Taylor solution for the blast wave, in the manner adopted by FWK00 for GRB970508.
Two other physical parameters that determine the evolution of the radiation are the fraction of total energy in relativistic electrons ($\varepsilon_{\rm{e}}$)
and in the post-shock magnetic field ($\varepsilon_{\rm{B}}$). 
In order to determine these four quantities, one requires the measurement of four spectral parameters, 
traditionally the three break frequencies and the flux normalisation. 
In the late phase, however, direct determination of the cooling frequency is difficult,
since the afterglow is not detectable at frequencies above radio bands. 
We therefore express the physical parameters as a function of the ratio $\varepsilon_{\rm{r}} \equiv \varepsilon_{\rm{e}}/\varepsilon_{\rm{B}}$; 
$\varepsilon_{\rm{r}} = 1$ would signify an equipartition of energy between the magnetic field and the relativistic particles.

Following Eq.~(5) of FWK00, we may then write the post-shock electron number density as
\begin{equation}
n = \varepsilon_{\rm{r}} \left( \frac{p-2}{p-1} \right)\frac{B_0^2}{8\pi\gamma_0 m_e c^2}
\label{eq:n_epsilon}
\end{equation}

We insert this in Eq.~(A12) of FWK00, and invert their Eqs.~(A10) to (A12) while making use of the relations (A6) to (A8). 
This yields the following set of expressions for the FWK00 model parameters:
\begin{eqnarray}
\gamma_0 & = & 88.3 \; d_{28}^{2/17}
               \left(\frac{\varepsilon_{\rm{r}}}{\eta_{10}}\right)^{2/17}
               \left(\frac{f_2^3}{f_3}\right)^{1/17} \nonumber \\
        & &    \cdot \left[(p-2)^2(p+2)^3\right]^{1/17}
               \left(\frac{\nu_{\rm{m}0,{\rm{GHz}}}}{\nu_{\rm{a}0,{\rm{GHz}}}}\right)^{3(p+4)/34}
	       F_{\rm{m}0,{\rm{mJy}}}^{1/17} \label{eq:gamma0} \\
r_0 & = & 1.407\times 10^{18} \;{\rm cm} \;\frac{d_{28}^{16/17}}{(1+z)}
               \left(\frac{\eta_{10}}{\varepsilon_{\rm{r}}}\right)^{1/17}
               \left(\frac{f_2^7}{f_3^8}\right)^{1/17} \nonumber \\
        & &    \cdot \left[\frac{(p+2)^7}{(p-2)}\right]^{1/17}
               \nu_{\rm{m}0,{\rm{GHz}}}^{(7p-6)/34}
               \nu_{\rm{a}0,{\rm{GHz}}}^{-7(p+4)/34}
	       F_{\rm{m}0,{\rm{mJy}}}^{8/17} \label{eq:r0} \\
B_0 & = & 4.58\times 10^{-2} \;{\rm G}\; \frac{(1+z)}{d_{28}^{4/17}}
               \left(\frac{\eta_{10}}{\varepsilon_{\rm{r}}}\right)^{4/17}
               \left(\frac{f_3}{f_2^3}\right)^{2/17} \nonumber \\
        & &    \cdot \left[(p-2)^4(p+2)^6\right]^{-1/17}
               \nu_{\rm{m}0,{\rm{GHz}}}^{(5-3p)/17}
               \nu_{\rm{a}0,{\rm{GHz}}}^{3(p+4)/17}
	       F_{\rm{m}0,{\rm{mJy}}}^{-2/17} \label{eq:B0} \\
n_i & = & \frac{0.69 \;{\rm{cm}}^{-3}\;}{1+X_{\rm{H}}} \frac{(1+z)^2}{d_{28}^{10/17}}
               \left(\frac{\varepsilon_{\rm{r}}}{\eta_{10}}\right)^{7/17}
               \left(\frac{f_3}{f_2^3}\right)^{5/17} \nonumber \\
        & &    \cdot \frac{1}{(p-1)}\left[\frac{(p-2)^7}{(p+2)^{15}}\right]^{1/17}
               \nu_{\rm{m}0,{\rm{GHz}}}^{(8-15p)/34}
               \nu_{\rm{a}0,{\rm{GHz}}}^{15(p+4)/34}
	       F_{\rm{m}0,{\rm{mJy}}}^{-5/17} \label{eq:ni}
\end{eqnarray}
In the above, $r_0$ is the radius of the blast wave, $B_0$ is the post-shock magnetic field, 
and $\gamma_0$ is the lower cutoff of the power-law distribution of relativistic electron Lorentz factors, all at the reference time $t_0$.
From these, the blast wave energy may be computed as:
\begin{eqnarray}
E_{\rm{ST}} & = & n_i m_p\left(\frac{1+z}{t_0}\right)^2 \left(\frac{r_0}{\xi}\right)^5 \label{eq:ESTdef} \\
           & = & \frac{8.576\times 10^{56} \;{\rm{erg}}\;}{(1+X_{\rm{H}})\xi^5 t_{0,{\rm{d}}}^2} 
                 \frac{d_{28}^{70/17}}{(1+z)}
                 \left(\frac{\varepsilon_{\rm{r}}}{\eta_{10}}\right)^{2/17}
                 \left(\frac{f_2^{20}}{f_3^{35}}\right)^{1/17} 
		 \frac{(p+2)}{(p-1)} \nonumber \\
          & &    \cdot \left[(p-2)^2(p+2)^3\right]^{1/17}
                 \nu_{\rm{m}0,{\rm{GHz}}}^{(10p-11)/17}
                 \nu_{\rm{a}0,{\rm{GHz}}}^{-10(p+4)/17}
	         F_{\rm{m}0,{\rm{mJy}}}^{35/17} \label{eq:EST} 
\end{eqnarray}
In Eq.~(\ref{eq:gamma0}) through (\ref{eq:EST}), $\nu_{\rm{a}0,{\rm{GHz}}}$ and $\nu_{\rm{m}0,{\rm{GHz}}}$ are the two break frequencies, in GHz units, 
at the reference time $t_0=t_{0,{\rm{d}}}$~days.  
The expressions assume that $\nu_{\rm{a}0,{\rm{GHz}}} > \nu_{\rm{m}0,{\rm{GHz}}}$, which is indeed the case in GRB030329 at these late times. 
$F_{\rm{m}0,{\rm{mJy}}}$ is the normalisation of the flux at an observing frequency $\nu \gg \nu_{\rm{a}},\nu_{\rm{m}}$, expressed as 
\begin{equation}
F_{\nu}(t=t_0) \;{\rm mJy}\; = F_{\rm{m}0,{\rm{mJy}}}\left(\frac{\nu}{\nu_{\rm{m}0}}\right)^{-(p-1)/2}
\end{equation}
$z$ is the redshift of the burst and $d_{28}$ the corresponding luminosity distance. 
The thickness of the post-shock emitting region at any time is assumed to be $r/\eta$, 
where $r$ is the radius of the blast wave and $\eta=10\eta_{10}$.
$f_2$ and $f_3$ are integrals over the synchrotron function defined in FWK00; both are functions of $p$. 
The quantity $\xi$, close to unity, is an equation of state-dependent normalisation factor for the blast wave radius (FWK00). 
$X_{\rm{H}}$ represents the mass fraction of hydrogen in the circumburst medium, and may nominally taken to be 0.75.

Evaluating spectral parameters from the fitted model, one finds, at a reference time $t_0 = 500\;{\rm{days}}$: 
$\nu_{\rm{m}0,{\rm{GHz}}} = 2.56\times 10^{-3}$, $\nu_{\rm{a}0,{\rm{GHz}}}=0.498$ and $F_{\rm{m}0,{\rm{mJy}}}=28.4$.  
Using $z=0.1685$ for GRB030329 and the fitted value of $p=2.1$, one then estimates
\begin{eqnarray}
E_{\rm{ST}} & = & 2.6\times 10^{51} \;{\rm{erg}}\; \left(\frac{\varepsilon_{\rm{r}}}{\eta_{10}}\right)^{0.12} \\
n_i & = & 0.35 \;{\rm{cm}}^{-3}\; \left(\frac{\varepsilon_{\rm{r}}}{\eta_{10}}\right)^{0.41}
\end{eqnarray}
The blast wave radius at 500~days works out to be $r_0 = 0.49 (\varepsilon_{\rm{r}}/\eta_{10})^{-0.06}$~pc.
The corresponding post-shock magnetic field is $B_0 = 0.036 (\varepsilon_{\rm{r}}/\eta_{10})^{-0.24}$~G,
and the lower cutoff of the electron Lorentz factor distribution at that time is 
$\gamma_0 = 5.5 (\varepsilon_{\rm{r}}/\eta_{10})^{0.12}$. 
These yield $\varepsilon_{\rm{e}} = \varepsilon_{\rm{r}} \varepsilon_{\rm{B}} = 0.085 (\varepsilon_{\rm{r}}/\eta_{10})^{0.24}$.


\section{Discussion}\label{section:discussion}

The data presented in this paper give an unprecedented view of the non-relativistic evolution phase of a GRB blast wave, 
because of the wide range covered in both frequency and time. 
This gives us the opportunity to compare the physical parameters that we have derived from the very late-time data 
with the physical parameters derived from the early-time data, when the blast wave was still extremely relativistic. 
From the emerging physical picture we put constraints on the emission from the counter jet, 
and we compare our findings with the results from VLBI measurements of the source size evolution.


\subsection{Relativistic versus Non-Relativistic}

In Section \ref{section:parameters} we calculated the total energy in the blast wave $E_{\rm{ST}}$ and the density of the circumburst medium $n_i$ 
as functions of the ratio $\varepsilon_{\rm{r}} \equiv \varepsilon_{\rm{e}}/\varepsilon_{\rm{B}}$, assuming that the blast wave was in its non-relativistic phase. 
These parameters are determined in \citet{vanderhorst2005:apj634} in the relativistic phase as functions of the cooling frequency, adopting $\nu_{\rm{c}}=10^{13}$~Hz. 
We calculate the energy and density in the relativistic phase as functions of $\varepsilon_{\rm{r}}$, 
a better way to compare the parameters in the two different phases. 
Since the values of $\varepsilon_{\rm{e}}=0.25 \nu_{\rm{c},13}^{1/4}$ and $\varepsilon_{\rm{B}}=0.49 \nu_{\rm{c},13}^{-5/4}$ 
in \citet{vanderhorst2005:apj634} indicate near-equipartition, the values for the density and energy do not differ much from the values obtained there: 
$n_{\rm{r}}=0.78\times\varepsilon_{\rm{r}}^{1/2}\,{\rm{cm}}^{-3}$, $E_{\rm{r,iso}}=2.6\times 10^{51}\times\varepsilon_{\rm{r}}^{1/6}\,\rm{erg}$ 
and $E_{\rm{r,cor}}=3.4\times 10^{50}\times\varepsilon_{\rm{r}}^{1/4}\,\rm{erg}$, the latter being the beaming-corrected energy. 
The cooling frequency can be found to be $\nu_{\rm{c}}=1.6\times 10^{13}\times\varepsilon_{\rm{r}}^{2/3}$~Hz, 
validating the assumptions in previous studies on $\nu_{\rm{c}}$ being $\sim 10^{13}$~Hz at 10~days.

We find that the density derived in the non-relativistic phase is a factor of two smaller than the density derived in the relativistic phase. 
The beaming-corrected energy $E_{\rm{r,cor}}$, however, is a factor of $\sim 7$ times smaller than the total energy $E_{\rm{ST}}$ in the blast wave 
derived from the non-relativistic evolution. 
\citet{frail2005:apj619} modeled the first year of observations with the VLA and ATCA, and found a total kinetic energy of $9.0\times 10^{50}$~ergs; 
using only the data before 64~days they derive an energy of $6.7\times 10^{50}$~ergs \citep[see also][]{berger2003:nature426}, 
a factor of $\sim 2$ larger than our $E_{\rm{r,cor}}$;
and using only the data after 50 days, to obtain an estimate for $E_{\rm{ST}}$, they find an energy of $7.8\times 10^{50}$~ergs, 
a factor of $\sim 4$ smaller than our value for $E_{\rm{ST}}$. 
In these models the range for the circumburst density is $\sim 1-3\,{\rm{cm}}^{-3}$, a bit larger than our values. 
\citet{granot2005:apj618} also determine the energy and density with their models, and find a similar value for the collimation corrected energy as we do. 
Their value for the density is, however, an order of magnitude larger, but they note that this can be attributed to the fact that the density depends strongly 
on the precise value of $\nu_{\rm{a}}$ and $\nu_{\rm{m}}$. 

Given the differences in the methods used by different authors and the uncertainties in the assumptions made to estimate these numbers, 
they can all be considered to be comparable. 
It is not possible to make any definite statements about significant differences in energies derived from the relativistic and non-relativistic phase. 
If, however, the somewhat larger estimate of the total energy in the non-relativistic phase is indeed true, 
then two possible explanations may be advanced for this:
either the beaming correction in the relativistic phase is too strong, giving a too small value for the beaming corrected energy; 
or $E_{\rm{ST}}$ is over-estimated because of non-isotropy in the emission coming from the blast wave in the non-relativistic phase. 
In the latter case it could be that the blast wave is not completely spherical yet, but still the evolution is well described by the Sedov-Taylor solution; 
or that the blast wave is spherical, but the emission is not coming from the blast wave isotropically. 
In both situations the value of $E_{\rm{ST}}$ that we derived would be an over-estimate of the true value. 

The energies that we and \citet{frail2005:apj619} derive indicate that our estimate of the relativistic beaming 
and the assumption that the blast wave becomes spherical at $t\simeq t_{\rm{NR}}$ are valid. 
The latter is important for testing the models that describe the lateral spreading of the collimated outflow after the jet-break time, 
when the Lorentz factor drops below the inverse of the half-opening angle of the jet. 
Some (semi-analytical) models \citep[e.g.][]{rhoads1999:apj525} assume a very rapid sideways expansion of the jet 
with a lateral expansion velocity of the order of the velocity of light, 
resulting in an exponential growth of the jet half-opening angle with radius. 
Hydrodynamical simulations, however, show a very modest degree of lateral expansion as long as the jet is relativistic 
\citep[for an extensive review, see][]{granot2007:RevMexAstro27}. 
In the latter model the outflow is still strongly collimated when the blast wave becomes non-relativistic, 
while the first model predicts that the blast wave is (almost) spherical at $t\simeq t_{\rm{NR}}$, 
which is favoured by our analysis, since there is no significant change in temporal slopes well after $t_{\rm{NR}}$.


\subsection{Counter Jet Emission}

It is expected that there are two collimated outflows formed at the collapse of a massive star: two jets pointed away from each other. 
When the counter jet becomes non-relativistic, the emission is no longer strongly beamed away from us. 
\citet{granot2003:apj593} predicted a re-brightening in the radio light curves when that occurs, 
although it would be difficult to detect in the case of GRB\,030329. 
\citet{li2004:apj614} calculated that the observed time at which the counter jet becomes non-relativistic is 5 times larger than $t_{\rm{NR}}$, 
because of light travel time effects. 
They claim that the observed flux coming from the counter jet at that time is comparable to the flux of the jet coming towards us at $t_{\rm{NR}}$. 
This results in a rapid increase in flux with a peak at $5\times t_{\rm{NR}}$, 
and after the peak the light curve declines steeper than before the rapid increase until it relaxes to the original light curve behaviour. 
\citet{li2004:apj614} predicted such a feature in the radio light curves of GRB\,030329 at $\sim 1.7$~years after the burst.  
From Figure \ref{fig:results}, however, it is clear that this is not observed up to 3~years after the burst 
\citep[which was also noted, with observations up to 2~years after the burst, by][]{pihlstrom2007:astroph07042085}. 

The apparent discrepancy between the predictions for the counter jet and the observed light curves can be explained by looking carefully at the assumptions. 
The calculations by \citet{li2004:apj614} are valid for those observing frequencies at which the light curve peaks 
when the blast wave is still ultra-relativistic and narrowly collimated. 
This means that it can only be applied to the light curves at 8.4~GHz and higher frequencies. 
Adopting our value for $t_{\rm{NR}}$ of 60-80~days, this would mean that the 8.4~GHz light curve 
has to rapidly increase to a peak of 1-2~mJy at 300-400~days ($5\times t_{\rm{NR}}$). 
This is not observed, although it is hard to make definite statements about this, since there are only observations up to 360~days. 

The light curves at 2.3~GHz and lower frequencies are not expected to show this kind of re-brightening, 
because they peak at or after $t_{\rm{NR}}$. 
This results in a flattening of the peak to a width of $2\times t_{\rm{NR}}$. 
Looking at the low frequency light curves this could be the case, 
although the scatter in the data is too large to make any quantitative statements about this. 

The best constraints on the counter jet could come from the 4.8~GHz light curve. 
Following \citet{li2004:apj614} the flux at 300-400~days has to increase to 4-6~mJy, while the measured flux at that time is $0.6-0.8$~mJy. 
This means that the observed flux coming from the counter jet is suppressed by a factor of $\sim 8$, while at 8.4~GHz the limit was a factor of $\sim 4$. 
The flux from the counter jet, however, is probably lower than that predicted by \citet{li2004:apj614}, 
because, especially at 4.8~GHz, the light curve peaks later than the jet-break time. 
This means that the two jets, which already have quite a large opening angle of $\sim 20^{\circ}$ to start with, 
cannot be treated as two narrowly collimated outflows anymore because of lateral spreading of the jet.

Concluding, we cannot say, from the light curves presented here, whether we have seen emission from the counter jet or not. 
The fact that we do not see a late-time re-brightening at high radio frequencies 
could be due to the fact that the outflow is not very narrowly collimated. 
A flattening of the peak of the light curves at low radio frequencies, caused by the emission coming from the counter jet, 
could be present, but the scatter in the data prevents us from drawing any firm conclusion regarding this.


\subsection{Source Size Evolution}

VLBI observations of GRB\,030329 make it possible to study the image size of the afterglow. 
The full set of VLBI measurements up to 806~days after the burst has been presented in \citet{pihlstrom2007:astroph07042085}. 
\citet{granot2005:apj618} have discussed the implications of the measured image sizes up to 83~days after the burst, 
which have mainly remained the same after including the latest measurements. 
They conclude that a homogeneous medium gives a better fit to the image size evolution than a stellar wind environment, 
although it is hard to rule out the latter due to the measurement uncertainties. 
Furthermore, they try to constrain the amount of lateral spreading, i.e. whether the lateral expansion velocity is of the order of the velocity of light, 
or that there is hardly any lateral spreading until the blast wave becomes sub- or non-relativistic. 
This is, however, hard to constrain, again because of the uncertainties in these source size measurements. 

An important parameter that comes out of the VLBI studies is $t_{\rm{NR}}$. 
\citet{granot2005:apj618} and \citet{pihlstrom2007:astroph07042085} claim that $t_{\rm{NR}}\sim 1$~year from looking at the evolution 
of the apparent expansion velocity, and from theoretical model fits to the source size evolution. 
The models in which there is rapid lateral expansion, their model 1, $t_{\rm{NR}}\sim 1$~year is indeed obtained, 
but the jet-break time in that case is found to be $\sim 1$~month instead of 10~days \citep[e.g., see Figure 4 of][]{pihlstrom2007:astroph07042085}. 
Since the jet-break time is quite well determined from the light curves, the value of $t_{\rm{NR}}$ in this model 1 is rather uncertain. 
The models in which there is no lateral expansion until the blast wave becomes non-relativistic, their model 2, 
the non-relativistic transition actually happens at 60-80~days after the burst; 
the blast wave then becomes spherical on a time scale of 1-3~years. 
So the fits of model 2 do indeed indicate a value of $t_{\rm{NR}}\sim 60-80$~days.

From the evolution of the average apparent expansion velocity the value of $t_{\rm{NR}}$ can also be deduced. 
The uncertainties and correction factors in deriving the average velocity are discussed thoroughly in \citet{pihlstrom2007:astroph07042085}. 
From their Figure 3 it seems that the transition happens at $\sim 100-200$ days, when the average apparent expansion velocity is $\sim 2$ times the speed of light. 
The uncertainties in the correction factors applied, however, make it hard to state that this is significantly higher 
than the value of $t_{\rm{NR}}$ that is deduced from the light curves. 
The issue that still remains, is that the value of $t_{\rm{NR}}$ that various authors have estimated from energy considerations 
is a factor of a few higher than 60-80~days \citep[e.g.][]{granot2005:apj618,vanderhorst2005:apj634}. 
This is, however, a very rough estimate, and a discrepancy within a small factor should not be regarded as disturbing. 
In conclusion, the value of $t_{\rm{NR}}$ determined from the VLBI source size measurements 
is not at significant odds with the value of $t_{\rm{NR}}$ derived from light curve studies.


\subsection{Low Frequency Array}

\begin{figure}
  \centering
  \includegraphics[width=\columnwidth]{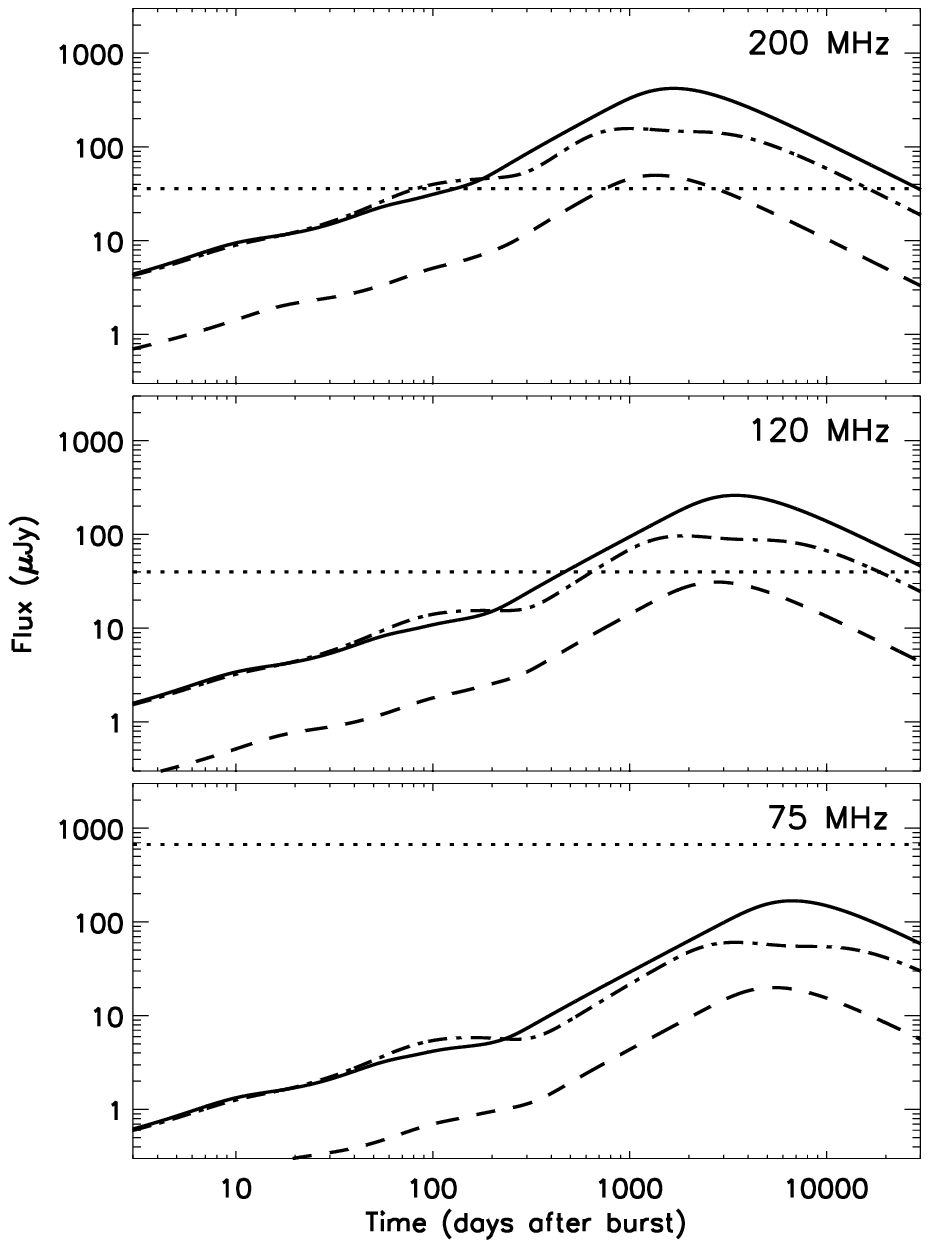}
    \caption{The predicted light curves of GRB\,030329 at three frequencies within the LOFAR observing range. 
        The solid line and dash-dotted line correspond to the two models shown in Figure \ref{fig:results}; 
	the dashed line shows the predicted light curve of GRB\,030329 when situated at a redshift of 1 instead of 0.16. 
	The horizontal dotted lines are the 1-$\sigma$ sensitivity limits after 4~hours of observing with a bandwidth of 4~MHz. 
	It shows that it is quite feasible to detect GRB afterglows similar in brightness to that of GRB\,030329 with LOFAR on timescales of months to decades; 
	and that fainter afterglows can also be detected after long integration times.}
      \label{fig:lofar}
\end{figure}

From our analysis of the broadband radio afterglow of GRB\,030329 we can calculate predicted light curves for future radio telescopes 
operating in the metre wavelength regime. 
As an example we explore the possibilities for the Low Frequency Array (LOFAR) to detect the GRB\,030329 afterglow and other GRB afterglows. 

LOFAR will be a major new multi-element, interferometric, imaging telescope designed for the 30-240~MHz frequency range. 
LOFAR will use an array of simple omni-directional antennas, whose electronic signals are digitised, 
transported to a central digital processor, and combined in software to emulate a conventional antenna. 
LOFAR will have unprecedented sensitivity and resolution at metre wavelengths, 
and will be the first of a new generation of radio telescopes to become fully operational, i.e. in early 2009. 
This sensitivity and resolution will give the GRB community the opportunity to study bright afterglows 
on even longer timescales than with observations at centimetre wavelengths. 
For a concise description of LOFAR and the Transients Key Project, in which GRBs are among the prime targets, 
see \citet{rottgering2006:astroph0610596} and \citet{fender2006:astroph0611298}.

The fact that our early-time model calculations from \citet{vanderhorst2005:apj634} gave such good predictions for the late-time behaviour, 
in particular at observing frequencies below 1~GHz, gives us confidence to extrapolate the modeling results of the radio afterglow of GRB\,030329 
to the LOFAR observing range, see Figure \ref{fig:lofar}. 
The predicted light curves show that GRB\,030329 will be observable in the high band of LOFAR (120-240~MHz), but not in the low band (30-80~MHz). 
The light curves are expected to peak in 2009, when LOFAR will be fully operational, and even later going down from 240 to 120~MHz; 
the expected peak flux is $\sim 0.3-0.4$~mJy in the LOFAR high band. 
We also calculated light curves for GRB\,030329 if it were situated at a redshift of 1 instead of 0.16. 
The resulting fainter afterglow can also be detected, although with longer integration times, i.e. on the order of a day instead of an hour.


\section{Conclusions}\label{section:conclusions}

We have presented a detailed study of the late-time radio afterglow of GRB\,030329. 
We have obtained measurements with the WSRT and GMRT, spanning a spectral range of 325~MHz-8.4~GHz and a temporal range of 268-1128~days after the burst. 
Combined with all the already published radio observations of this afterglow, from WSRT, GMRT and other large radio telescopes, 
we have studied the physics of the blast wave in the non-relativistic phase and compared these results with those from studies of the relativistic phase. 
The well-sampled late-time light curves made it possible to determine the index of the electron energy distribution accurately, 
and to confirm that the circumburst medium is homogeneous.

The energy of the blast wave and density of the circumburst medium, determined from the non-relativistic evolution of the blast wave, 
are comparable to findings by several studies based on earlier time observations. 
We have shown that the blast wave is spherical, or almost spherical at least, at the time that the blast wave becomes non-relativistic: $t_{\rm{NR}}\sim 80$~days. 
In contrast with some predictions, a radio re-brightening due to the counter jet becoming non-relativistic, is not observed. 
The existence of a counter jet cannot be ruled out, since it is possible that the peaks of the light curves at low radio frequencies are flattened 
due to this counter jet. 
We have also shown that the high-resolution VLBI measurements of the afterglow image size are in agreement with our light curve studies. 
In particular, the value of $t_{\rm{NR}}$ derived from modeling the image size evolution does not differ from our findings significantly 
within the measurement and modeling uncertainties.

The afterglow of GRB\,030329 will be followed-up at radio frequencies for many years to come. 
The current brightness and the fact that the flux drops logarithmically, 
make it possible to continue studying this afterglow for the next decade at least. 
VLBI capabilities will increase significantly in the coming years, 
so the image size evolution can also be observed. 
A new generation of radio telescopes will also be able to observe this afterglow. 
The first one of this new generation to come on-line is LOFAR, which will be able to detect GRB\,030329 at metre wavelengths. 
Besides this extremely bright afterglow, LOFAR will also be able to study many afterglows at frequencies 
and on timescales that so far have been unexplored.


\begin{acknowledgements}
We thank the referee for constructive comments. 
We greatly appreciate the support from the WSRT and GMRT staff in their help with scheduling these observations. 
The Westerbork Synthesis Radio Telescope is operated by ASTRON (Netherlands Foundation for Research in Astronomy) 
with support from the Netherlands Foundation for Scientific Research (NWO). 
The Giant Metrewave Radio Telescope is operated by the National Center for Radio Astrophysics of the Tata Institute of Fundamental Research. 
RAMJW gratefully acknowledges the support of NWO under grant 639.043.302. ER gratefully acknowledges support from PPARC. 
DB thanks the Astronomical Institute 'Anton Pannekoek' for hospitality during a part of this work.
\end{acknowledgements}

\bibliographystyle{aa}
\bibliography{references}

\end{document}